\documentclass[aps,prb,twocolumn,superscriptaddress,floatfix,citeautoscript,preprintnumbers]{revtex4-2}

\usepackage{graphicx,color}
\usepackage{amssymb}
\usepackage{amsmath}
\usepackage{epstopdf}
\usepackage[normalem]{ulem}
\usepackage{comment}
\usepackage{breakcites}
\usepackage[breaklinks=true]{hyperref}
\usepackage[caption=false]{subfig}
\usepackage{enumitem}
\usepackage{soul}
\usepackage{bbold,dsfont}

\begin{document}


\title{Tensor Network Algorithms:  a Route Map}

\author{Mari Carmen Ba\~nuls} 
\affiliation{
Max-Planck-Institut f\"ur Quantenoptik, Hans-Kopfermann-Str.1, Garching, Germany, D-85748; email: mari.banuls@mpq.mpg.de}
\affiliation{
Munich Center for Quantum Science and Technology (MCQST), Schellingstr. 4, Munich, Germany, D-80799}

\begin{abstract}
Tensor networks provide extremely powerful tools for the study of complex classical and quantum many-body problems.
Over the last two decades, the increment in the number of techniques and applications has been relentless, 
and especially the last ten years have seen an explosion of new ideas and results that may 
be overwhelming for the newcomer.
This short review introduces the basic ideas, the best established methods and some of the most significant algorithmic developments
that are expanding the boundaries of the tensor network potential.
The goal is to help the reader not only appreciate the many possibilities offered by tensor networks, but also find their way through  
state-of-the-art codes, their applicability and 
some avenues of ongoing progress.
\end{abstract}

\maketitle

\tableofcontents

\section{INTRODUCTION}

Numerical methods are an essential tool to tackle quantum many-body systems, 
most of which lack analytical solutions.
For these problems, though, 
 the dimensionality---and with it the computational complexity---grows
exponentially with the system size.
This limits the applicability
of exact numerical calculations and calls for
the development of numerical methods 
that can efficiently deal 
with, at least, the most relevant physical questions.
Introduced to the field in the nineties,
tensor network (TN) techniques aim to cover this need and
have become by now, together with exact diagonalization and Quantum Monte Carlo methods, a key instrument
in the numerical study of quantum many-body problems. 

{TN have been discovered independently in different disciplines. 
First uncovered in statistical physics by Baxter~\cite{Baxter1968dimers},}
in the field of {quantum many-body physics} their ancestry  can be traced back to the 
first valence bond solid (VBS) proposed by Affleck, Kennedy, Lieb and Tasaki~\cite{Affleck1987}
as exact ground state of a short-range spin chain 
{---see~\cite{Okunishi2021review} for a review and historical perspective.}
Kl\"umper et al.~\cite{Kluemper1993} later extended the AKLT proposal for a larger set of models, 
and also introduced the term \emph{matrix product} to designate these states.
The construction was generalized and formalized mathematically by Fannes, Nachtergaele and Werner~\cite{Fannes1992}
in the finitely correlated states for infinite spin chains.

Around the same time the density matrix renormalization group (DMRG), a new algorithm proposed by White~\cite{White1992}, 
was revealing an amazing power to capture the ground state of large quantum spin chains with only modest numerical effort.
Shortly afterwards, \"Ostlund and Rommer~\cite{Oestlund1995} identified the fixed point of the infinite DMRG algorithm with precisely such matrix product states,
and Dukelsky et al.~\cite{Dukelsky1998} pointed out the connection between DMRG and a variational search over these states.
Furthermore Nishino and Okunishi~\cite{NishinoOkunishi1995,NishinoOkunishi1996ctmrg} unified DMRG with Baxter's corner transfer matrix approach 
for two dimensional classical models.
And these insights inspired further generalizations of the original algorithm~\cite{Nishino1998threeD}.

DMRG was applied to multiple scenarios and fast became a method of choice to study 
the static properties of quantum spin systems in low spatial dimension~\cite{Hallberg2003dmrg,Schollwoeck2005}.
Yet a whole new perspective was gained thanks to quantum information concepts. Understanding in terms of entanglement the 
matrix product ansatz~\cite{Vidal2003} and the DMRG algorithm~\cite{Verstraete2004}, and reformulating the latter fully in terms of matrix product states (MPS)~\cite{McCulloch2007} 
opened up the possibilities for improvements and jumpstarted the \emph{tensor network} field. 
In particular, algorithms for real time evolution~\cite{Vidal2004,Daley2004,White2004real} and finite temperature~\cite{Verstraete2004a,Zwolak2004,Feiguin2005finiteT} 
with matrix product states, as well as a generalization to higher dimensions~\cite{Verstraete2004b} were proposed soon afterwards, revealing the potential of the tensor network picture.

Nowadays, TN algorithms are among the standard numerical methods for strongly correlated low-dimensional 
quantum systems. 
Most commonly used are the original methods from the early 2000s, which continuously find new applications.
But the TN language continues to be 
 exploited to provide, not only deeper mathematical understanding
of the ansatz~\cite{Cirac2021rmp}, but also new numerical techniques.

The variety of TN applications that have bloomed over the last decade and produced state-of-the-art results
is too vast to do justice to it in these pages. 
Thus, the focus of this article is the general framework of TN algorithms, with a stress on a few selected advances in
the field that are important for cutting-edge applications. 
The details of the algorithms are not explicitly shown; the interested readers 
 are encouraged to refer to the many excellent reviews in the literature,
such as~\cite{Verstraete2008,Schollwoeck2011,Orus2014annphys,Bridgeman2017,Ran2020tncontr,Silvi2019tn}, to name only a few.

\section{BASIC CONCEPTS}
\label{sec:basic}

\begin{center}
\begin{figure}[t]
\includegraphics[width=.8\textwidth]{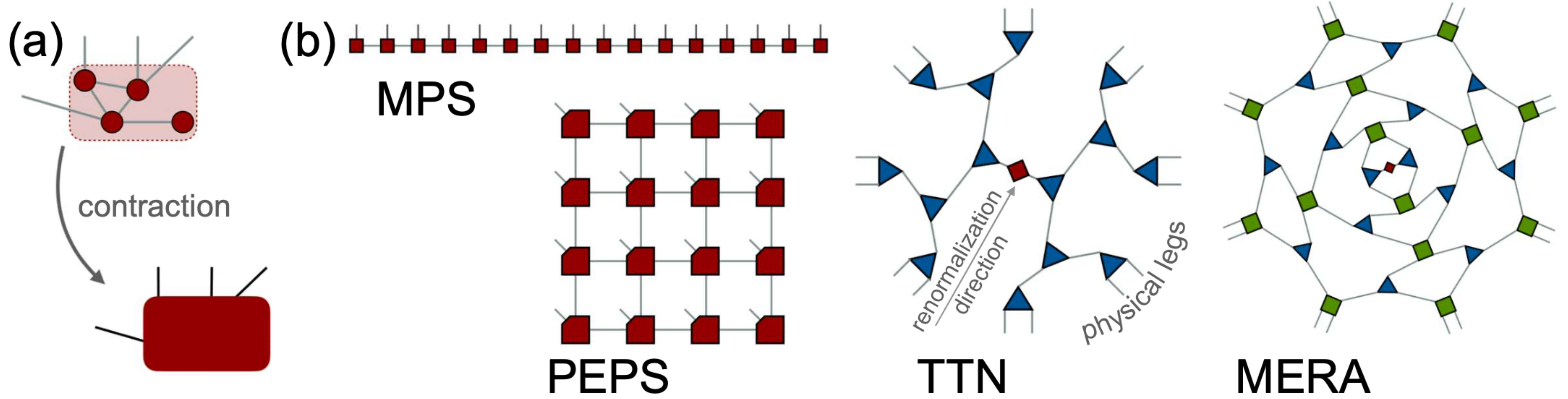}
\caption{Graphical representation of tensors: (\textit{a}) Example of a TN formed by four tensors; when contracted, a $4$-rank tensor is obtained; (\textit{b}) graphical representation of a TNS in each of the main families, for a system of 16 sites (note: triangles are commonly used to indicate isometries).
}
\label{fig1}
\end{figure}
\end{center}

A tensor, the basic object, is simply a multidimensional array.
The graphical representation of TN, illustrated in fig~\ref{fig1}, provides a practical language to describe their algorithms and properties.
For instance, a $k$-rank tensor, an object with $k$ indices, is depicted as a geometrical shape with $k$ legs (e.g. a  matrix would have two legs).
A contraction between two tensors---such as a matrix-vector product---is represented by joining the contracted indices.
In general, a tensor network is a set of such interconnected tensors, resulting in a
 rank determined by the number of open legs (see  fig~\ref{fig1}a).

\onecolumngrid

\twocolumngrid

\subsection{Tensor network states}
In particular, a \emph{tensor network state} (TNS) 
encodes all coefficients (in a given basis) of a quantum many-body state
in such a diagram, with as many open legs as constituents in the system.
Each dangling leg corresponds to the (finite) \emph{physical} dimension of one site,
while contracted legs correspond to  \emph{virtual} or \emph{bond dimensions}.

TNS families are defined by graphs with different connectivities.
For the families of interest, the number of parameters, proportional to the number of tensors, grows polynomially with the system size.
This represents a drastic reduction with respect to the exponentially large dimension of the Hilbert space.
But the aim of TNS is to capture physical states, which happen to explore only a small fraction of all possible quantum states, 
mainly characterized by their low entanglement. 
In particular, ground and thermal equilibrium states of local Hamiltonians fulfill an entanglement area law~\cite{Eisert2010area}:
the entanglement between a certain subsystem and the rest scales with the size of the boundary between both parts (or with small corrections thereof), 
instead of with the size of the bulk of the subsystems, as is the case for most states in the Hilbert space.
A rigorous proof of the area law scaling exists for gapped one-dimensional local Hamiltonians~\cite{Hastings2007} and for 
thermal equilibrium states in any dimension~\cite{Wolf08arealaw},
whereas critical ground states can display small (logarithmic) corrections~\cite{Calabrese2004,Wolf2005fermions}.

The following are the most widely used TNS families (see their diagrams in fig.~\ref{fig1}b).

\begin{enumerate}
\item MPS have a one-dimensional structure, with one tensor per lattice site~\cite{PerezGarcia2007}. 
Each tensor owns one open index corresponding to the physical dimension of the site, and two virtual legs connected to the neighboring sites (for open boundaries, the edge tensors only connect to one neighbor).
The tensor for site $k$ has components $A[k]^{i}_{\alpha\, \beta}$, where $i$ takes values over the physical dimension (typically denoted $d_k$),
and $\alpha$ and $\beta$ respectively take values over the left and right virtual dimensions of the tensor (denoted $D_{l}$ and $D_{r}$)---equivalently, each $A[k]^{i}$ is a $D_l\times D_r$ matrix.
More explicitly, for a system of $N$ sites, all with physical dimension $d$, the state can be written
\begin{equation}
|\Psi\rangle=\sum_{i_1,i_2,\cdots i_N=1}^{d} \mathrm{tr}\left ( A[1]^{i_1}A[2]^{i_2}\cdots A[N]^{i_N}\right)|i_1 i_2 \cdots i_N\rangle.
\label{eq:mps}
\end{equation}
MPS satisfy an entanglement area law: the half-chain entanglement of an MPS with maximal bond dimension $D$ is upper-bounded by $S=2 \log D$.
Furthermore, they hold exponentially decaying correlations, can be prepared  and contracted efficiently, and essentially correspond to ground states of local one-dimensional 
gapped Hamiltonians~\cite{Schuch2010peps}.
\item PEPS are the natural generalization of MPS to arbitrary graphs, where they can be 
defined with one tensor (with a physical leg) per vertex and connections according to the graph edges~\cite{Verstraete2004b}. 
They can be expressed analogously to Eq.~\ref{eq:mps}, replacing the trace by a contraction over all connections.
PEPS fulfill the area law in higher dimensions, and are much more complex objects than MPS. 
For instance, they cannot---in the general case---be prepared or contracted efficiently~\cite{Schuch2007complexity} and, even with small bond dimension, they can support critical correlations~\cite{Verstraete2006a}.
\item TTN correspond to tree graphs. Usually---but not always---they have physical indices in the leave nodes~\cite{Shi2006}, connected to tensors with only virtual indices  at higher levels (see fig.~\ref{fig1}b), {which can correspond} to a renormalization direction~\cite{Cirac2009rg}. 
As MPS, TTN are loop-free and can be contracted efficiently, but they violate the one-dimensional area law, and can hold power-law decaying correlations when averaging over spatial positions~\cite{Silvi2010tree}.
TTN can be used also for higher dimensional systems~\cite{Tagliacozzo2009ttn2D}. 
\item MERA implement a more complex renormalization of the physical degrees of freedom~\cite{Vidal2007,Vidal2008,Evenbly2009}, in which layers of unitary transformations (called disentanglers) that
remove short range correlations are alternated with layers of isometries that perform the renormalization step. 
This results in a TN with cycles in which,
thanks to the unitarity properties of the tensors, local expectation values can be computed efficiently. 
Scale invariant MERA can describe quantum critical ground states in one dimension~\cite{Pfeifer2009qcMERA,Montangero2009qcMERA}, where they support 
 logarithmic corrections to the area law.
However, in two dimensions they are proven to be a subset of PEPS~\cite{Barthel2010mera}, and thus satisfy the area law.
However, a generalization called branching MERA~\cite{Evenbly2014branchingMERA} exists that can support up to volume-law entanglement in more 
than one spatial dimension.
\end{enumerate}

Any tensor network has a so-called gauge freedom, since inserting 
the product of a matrix and its inverse $X X^{-1}$ in between any contracted pair of indices (i.e. in a connected leg) leaves the whole TN invariant, but 
allows redefining pairs of neighboring tensors.
 For loop-free TNS, in which cutting a bond splits the network in two, it is possible to define a canonical form, 
in which the basis for the virtual index is chosen to be the Schmidt basis 
for the bipartition corresponding to the bond, explicitly encoding the corresponding entanglement~\cite{PerezGarcia2007}.
Besides being fundamental to characterize the properties of a TNS family,
this canonical form gives rise to more stable and efficient numerical algorithms.
%

The families above can be defined for finite-size systems with site-dependent tensors, but it is also possible
to consider directly the thermodynamic limit, in which one (or a few) tensors are repeated infinitely many times,
to produce a translationally invariant (or periodic in space) structure.
In the case of MPS, the translationally invariant ansatz is called uniform MPS (uMPS). 
In infinite PEPS (iPEPS), a periodic iteration of a finite unit cell is most commonly used in practice, while the translationally invariant version is fundamental for the formal results~\cite{Cirac2021rmp}.
This allows targeting  bulk properties directly, without finite-size extrapolations, 
or, in the case of MERA, capturing the scale invariance of critical systems~\cite{Pfeifer2009qcMERA,Montangero2009qcMERA}.

These families can also describe mixed states. 
The simplest approach is to postulate the TNS ansatz in a given tensor product basis of the 
vector space of operators, with simply doubled physical legs.
In particular, in the MPS and PEPS cases the resulting structures are called MPO~\cite{Verstraete2004a,Zwolak2004,Pirvu2010a} and PEPO.
But if the ansatz is to describe a physical state, it must be positive semidefinite, a global property that cannot be assessed at the level of the local 
tensors. 
An alternative is to consider the TNS describing a purification, i.e. a pure state of the system plus an ancilla, such that 
tracing out the latter results in the desired mixed state. 
In the MPS and PEPS case, this yields a locally purified form, a TN with the same structure, where local tensors have double physical indices, and internal structure
granting positivity~\cite{Verstraete2004a}. This is more restrictive and potentially less efficient than the generic ansatz~\cite{delasCuevas2013},
but can be used in practice in numerical algorithms. 

\subsection{Fundamental primitives}
\label{subsec:primitives}

Virtually all TN algorithms rest on two basic 
blocks: contracting (part of) the tensor network, 
and locally updating the tensors.
Together with the approximation of (parts of a) TN by tensors with truncated dimensions, they
can be considered the fundamental primitives on which 
more or less sophisticated higher-level algorithms are built.

\subsubsection{Contracting TN}

A ubiquitous problem in TN algorithms is contracting a tensor network.  This means explicitly evaluating the products and sums of tensor components 
indicated by the connections, to result in a tensor with dimensionality corresponding to the indices that remain open (see fig.\ref{fig1}a).
For instance, for classical statistical models, partition functions and expectation values of local observables can be written 
as \emph{closed} TN (without open indices).
For TNS representing quantum states, norms and local expectation values are also closed tensor networks,  
 while reduced density matrices appear as smaller tensor networks with operator indices.
Two aspects of the contraction affect the implementation and performance of the algorithm.

\begin{enumerate}
\item{Contraction order.}
In general, the computational cost of contracting a series of tensors with each other depends on the order in which operations are applied.
For the regular networks that appear in the most common TNS algorithms, the number of possibilities is small, and the optimal sequence 
(which minimizes the computational cost) is known.
But in the general case, finding the optimal contraction order is a NP-complete problem, for which some heuristic algorithms exist
\cite{Pfeifer2014optimal,Gray2021hyperoptimized}. 
\item{Computational cost.}
If a contraction order exists whose computational cost grows only polynomially with the size of the network, we say that the TN can be contracted exactly.
Such is the case with TN that do not contain loops, for instance the networks corresponding to expectation values of
 multi-point correlators in MPS and TTN.
 Also for TNS with some unitary properties there are contractible quantities, for instance the norm or few-point correlators 
 evaluated in MERA.
The exact contraction of an arbitrary tensor network is however a $\# P$-complete problem~\cite{Schuch2007complexity}.
Thus, most algorithms involving TN in more than one dimension need to approximate the contractions, which is referred to as \emph{truncation} (\ref{subsubsec:truncation}).
\end{enumerate}

\subsubsection{Tensor update}
Many algorithms work by holding a TN description of the quantity of interest and iteratively improving it until some predefined level 
of convergence is attained.
The improvement proceeds by local changes, or updates, in which one (or few) tensors are modified in order to optimize the relevant cost function. 
Typically, the latter depends on all tensors in the network but only one is allowed to vary in each update step, while keeping the others fixed,
hence turning the problem into a local one.
A related concept is thus the \emph{environment of a tensor}, the part of a TN
that is complementary to the tensor being modified. 
This appears in the local cost function and needs to be evaluated by a (in many cases approximate) contraction,
in order to determine the proper update for the local tensor.

\subsubsection{Truncation}
\label{subsubsec:truncation}

Truncating a TN means reducing (some of) the dimensions of its tensors, ideally in such way that the global result does not change. 
A truncation can be part of an approximation strategy (e.g. for a PEPS, see~\ref{subsec:peps}), where it is used to control the dimension of a partial TN contraction.
In the context of quantum states represented as TNS, truncating typically means 
finding tensors to approximate the state within a given family. For instance after acting with an operation that, if applied exactly, 
would increase the tensor dimensions, as is often the case of time evolution or non-local operators (e.g.~\ref{subsec:tebd}).
And, more generally, truncating may refer to approximating a certain state with a TNS of fixed bond dimension.

In any truncation step, a decision is made as to which degrees of freedom to keep and which to discard. 
In a TNS, the fixed bond dimension upper-bounds the
amount of correlations the state can hold, and thus the truncation step in most algorithms can be related to entanglement properties.

\subsection{Classic algorithms}
\label{subsec:classic}

Numerous TN algorithms have been introduced in the last years, yet there are a few well-established methods
that are used to obtain state-of-the-art results in quantum many-body problems.
Many of them (specially for one-dimensional problems) are available as open-source implementations (see Related Resources at the end), 
making it possible to benefit from the numerical power of TN methods 
without the need to dive into implementation details. 
Furthermore, they are not difficult to implement, and can be easily adapted to solve other problems, beyond the ones they were originally designed for.
They constitute the true workhorses of TNS numerical results.


\begin{center}
\begin{figure*}[t]
\includegraphics[width=.8\textwidth]{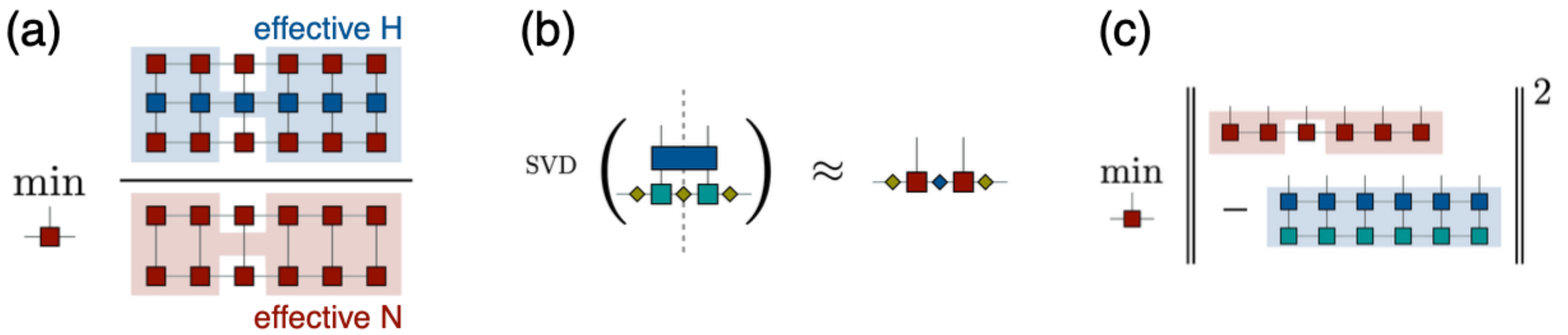}
\caption{Graphical expression of the local problems solved by the classic algorithms: (\textit{a}) variational optimization for a single tensor in DMRG; (\textit{b}) update of the local pair of tensors in TEBD (diamond-shaped tensors represent the Schmidt values, explicit in the canonical form); (\textit{c}) local optimization in tMPS.
}
\label{fig2}
\end{figure*}
\end{center}

\subsubsection{Variational optimization of MPS}
\label{subsec:variational}

One of the most powerful strategies in the TNS toolbox is the variational optimization of the ansatz with respect 
to a given cost function, the
paradigmatic example being the DMRG algorithm~\cite{White1992}.
This can be essentially understood as an application of the variational principle in which an ansatz for the ground state 
is obtained minimizing the energy for a quantum many-body Hamiltonian 
over the set of MPS with fixed bond dimension $D$~\cite{Verstraete2004,Schollwoeck2011},
\begin{equation}
| \Phi_{\mathrm{GS}}^{(D)} \rangle=\mathrm{argmin}_{|\Psi_{D}\rangle} \frac{\langle \Psi_D|H|\Psi_D\rangle}{\langle \Psi_D|\Psi_D\rangle}.
\label{eq:variational}
\end{equation}
The problem is tackled in an iterative manner, a single tensor being minimized at each step while the rest are kept constant.~\footnote{This corresponds to the single site algorithm, most natural in the TN framework. Some modifications can be made to connect to the classic two-site DMRG {[see~\cite{Schollwoeck2011,Okunishi2021review} for details on DMRG variants and their historical development]}.}
While not strictly necessary, the implementation of the original method is greatly simplified by writing the Hamiltonian as a MPO~\cite{McCulloch2007}.
This can be done exactly for short-range one-dimensional Hamiltonians~\cite{Pirvu2010a}, and approximation schemes exist for long-range interactions  
[e.g.~\cite{Hubig2017mpo}]. 
In this form, the local cost function can be written as the ratio of two tensor networks (see fig.~\ref{fig2}a) that can be contracted efficiently with a cost that, for $N$ sites, only scales as 
$O(N D^3)$. 
The local problem has thus the form of a Rayleigh-Ritz quotient, and can be solved exactly using a standard eigensolver. 
The procedure is iterated, sequentially optimizing each tensor in the ansatz, and repeatedly sweeping back and forth over the whole chain until 
a predetermined convergence criterion (usually convergence of the energy value within certain precision) has been reached.
Further gain in efficiency is possible if tensors are always kept in canonical form and intermediate calculations are stored in memory.
Because the optimum of each local problem can be found exactly, the algorithm is guaranteed to lower the energy monotonically, and thus to converge 
(even though this might be to a local minimum).

The infinite DMRG (iDMRG) algorithm directly targets systems in the thermodynamic limit~\cite{Schollwoeck2011}, and can also be expressed in similar terms.
In that case, instead of sweeping back and forth, at each step a unit cell of tensors is inserted and optimized in the middle of the chain,
and the procedure is iterated until a fixed point has been reached.

The most natural scenario for the algorithm is the search for the ground state of a local one-dimensional Hamiltonian. 
{The method is extremely competitive even for critical systems (for which the MPS can only approximate the correlations), thanks to finite-size and finite-entanglement~\cite{Pollmann2009finiteD,Pirvu2012finiteD} scaling,  and } 
has been successfully used for long-range interactions 
and problems in larger dimensions (see sec.~\ref{subsec:peps}). 
The efficiency and robustness of the method make it one of the most powerful numerical methods available to 
solve quantum many-body problems.
Additionally, it can be applied to any variational optimization problem in which the cost function is
expressed in terms of an effective Hamiltonian with MPO structure [e.g.~\cite{Cui2015open}].

\subsubsection{Evolving MPS: TEBD, tMPS}
\label{subsec:tebd}

The Time Evolved Block Decimation (TEBD) algorithm~\cite{Vidal2003,Vidal2004} 
is arguably the simplest to implement, yet one of the most versatile methods in the TN toolbox.
The strategy was originally proposed for simulating the evolution of an MPS under a quantum circuit,
which can be written as a sequence of two-body, nearest-neighbor unitary gates.
Since each gate can increase the entanglement, 
its exact action on an MPS generally results in a larger bond dimension.  
Maintaining an efficient description of the state thus requires an approximation step that reduces (truncates) the bond dimension after the application of each gate.
The TEBD strategy proceeds via a local update, involving only the directly affected tensors, and corresponds to 
minimizing the distance between the transformed and updated states under the condition that
all the remaining tensors are kept invariant.
Exploiting the canonical form of the MPS, this can be achieved by a singular value decomposition of a single tensor,
 obtained when contracting together the gate and the local MPS tensors, including their environment, which encodes the state of the rest of the system
(see fig.~\ref{fig2}b).
In the TEBD truncation step, only singular values above a certain threshold are kept, and the discarded weight gives a measure of the error.

For a one-dimensional nearest-neighbour Hamiltonian, the time evolution operator can be approximated,  using a Trotter-Suzuki expansion,  
as a sequence of such two-body gates of the form $\exp(-i \delta h_{i})$, where $h_i$ is a two-body term and $\delta$ a short time step.
The method can thus be used to simulate the dynamics 
of an MPS with cost that scales as $O(D^3)$, for bond dimension $D$.
The scheme can be adapted for other (finite-range) Hamiltonians, although the cost increases steeply with the interaction range. 

As an alternative to the local truncation, it is also possible to vary all tensors in order to minimize the distance to the exact state after one or more gates~\cite{Verstraete2006a}.
In this strategy, called tMPS,\footnote{{Notice that the term is used loosely in the literature, sometimes interchanged with tDMRG---see~\cite{Schollwoeck2011} for the details.}}
tensors are optimized sequentially as in the variational method~\ref{subsec:variational}, by solving a local problem that, in this case,
reduces to a  system of linear equations, also with cost $O(D^3)$ (see fig.~\ref{fig2}c). 
In this way once can apply onto the MPS vector any MPO operator, 
 in particular, a step of the Trotterized time-evolution. 
 The cost of such MPO representation also increases with the range of the Hamiltonian, but
long range interactions can be treated with help of an approximation scheme~\cite{Zaletel2015long}. 

These methods are very efficient and extraordinarily versatile.
Starting from an arbitrary state, the ground state can be approached by imaginary (or Euclidean) time evolution,
which effectively projects the state onto its lowest energy component,
and can be applied with the same algorithm~\cite{Vidal2004}, only using non-unitary terms $\exp(\delta h_{i})$.
Also thermal equilibrium states can be approximated using this technique~\cite{Verstraete2004a,Zwolak2004,Feiguin2005finiteT}, 
by writing a purification of the Gibbs ensemble (namely the thermofield state) as the evolution of a maximally entangled initial state in imaginary time given 
by the inverse temperature, $|\Psi\rangle\propto e^{-\beta H/2} \sum_n |n\rangle |n\rangle$ (where $n$ label a basis of the system Hilbert space).
 And by treating the mixed state as a vector in operator space, the same basic method can be used to simulate real time evolution of
open systems under master equations~\cite{Verstraete2004a,Zwolak2004}.
 Imaginary time evolution of pure states can also be used to produce a sample of minimally entangled typical
thermal states (METTS)~\cite{White2009metts} that reproduce thermal properties.
These are only a few examples:
more generally, the tMPS strategy approximates the action of any linear operator written as an MPO onto an MPS.
This allows reformulating most linear algebra algorithms as approximate versions in the framework of MPS [e.g.~\cite{GarciaRipoll2006,Huckle2012subspace}]. 
And TEBD and tMPS algorithms can also be applied to translationally invariant (or periodic) MPS, working directly in the thermodynamic limit~\cite{Vidal2007infinite,Verstraete2008}.

Even though the technique to treat all the scenarios named above is almost identical, the entanglement in each of them, and 
thus the performance of the method, widely differs.
While thermal equilibrium states satisfy an area law~\cite{Wolf08arealaw,Dubail2017mpo} and admit efficient TNS approximations 
for local Hamiltonians~\cite{Hastings2006a,Molnar2015}, 
real time evolution of a far-from-equilibrium state can give rise to linear growth of entanglement, in which case
approximating the resulting state with an MPS would require the bond dimension to grow exponentially with the total time~\cite{Calabrese2005,Osborne2006,Schuch2008}.
For this reason, while MPS methods are extremely useful to study 
dynamics close to equilibrium, or for moderate times~\cite{Paeckel2019tevol}, 
they suffer a fundamental limitation for genuinely out-of-equilibrium scenarios.

\section{ADVANCED TNS METHODS} 
\label{sec:advanced}

Even though the algorithms described in the previous section can treat a large number of problems, 
some more advanced techniques, developed mostly in the last decade and not yet freely available, 
are necessary to fully exploit the power of TNS.

\subsection{Higher dimensions}
\label{subsec:peps}


Despite the resounding success of the one-dimensional applications of TNS, applications of higher-dimensional ansatzes 
remain much less common.
However, in the last years, the situation has started to change, thanks to 
to a number of algorithmic developments and an intense effort of the community.

Treating two- and higher-dimensional problems has always been a coveted target of these numerical methods and,
since the early times of DMRG, the possibility was recognized of applying the technique to two-dimensional 
quantum~\cite{Stoudenmire2012}  
and three-dimensional classical problems~\cite{Nishino1998threeD}.
MPS form a complete family, and can be used as an ansatz for any problem, 
in particular  in larger dimensions.
For two-dimensional quantum states, the MPS ansatz can be wrapped around the lattice. This is usually done in a 
zig-zag or snake form, but other choices are possible~\cite{Cataldi2021hilbert}. 
The resulting representation of the Hamiltonian as MPO is more expensive (since some short-range terms get mapped onto longer-range ones),
and a larger bond dimension is required to reach the desired convergence: 
since cutting a single bond partitions the state in two, to accommodate the entanglement of a state that satisfies an area law  
the bond dimension needs to grow exponentially with one of the dimensions of the system.
Highly accurate computations are still obtained from systems of limited size,
often exploiting a long-cylinder geometry and careful finite size extrapolations [e.g.~\cite{Yan2011spinliquid,Depenbrock2012spinliquid}].

With built-in area law, PEPS are a more suitable TNS ansatz, which supports good approximations 
for equilibrium states of local Hamiltonians~\cite{Hastings2007,Molnar2015}, 
and allows variational and evolution strategies as described in 
section~\ref{subsec:classic}~\cite{Verstraete2004b,Murg2007peps}.
They can also be used directly in the thermodynamic limit, in which case they are called iPEPS, and are parametrized by a unit cell 
with a finite number (which could be as small as one) of tensors~\cite{Jordan2008ipeps}. 
Nevertheless, numerical algorithms with PEPS are considerably more involved, and have higher computational cost in terms of the tensor dimensions.


\begin{center}
\begin{figure*}[t]
\includegraphics[width=.8\textwidth]{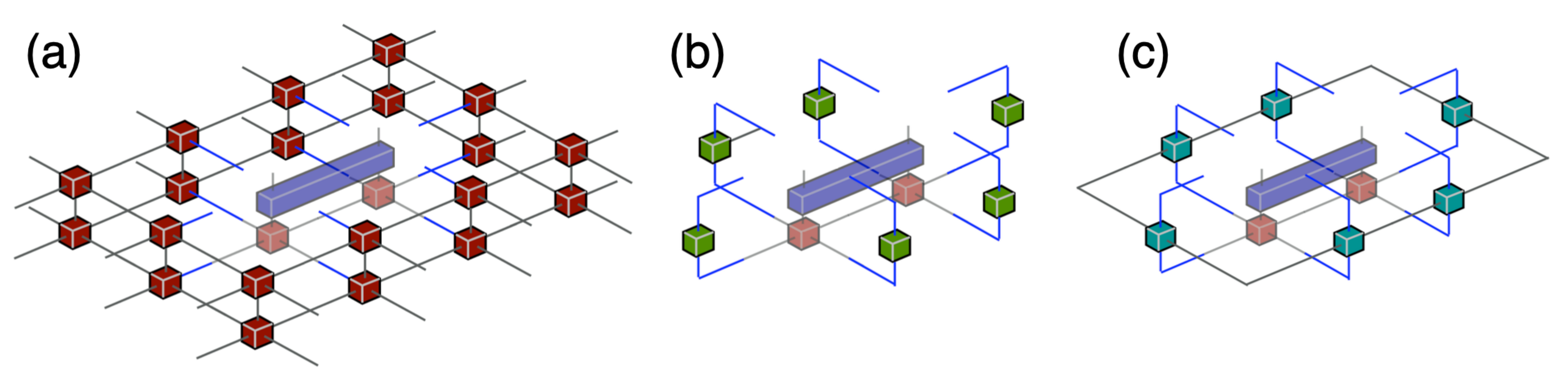}
\caption{Approximation of the environment tensor in PEPS: (\textit{a}) environment (in solid colors) of a pair of tensors on which a nearest-neighbor gate is applied (group shown in lighter shade); (\textit{b}) in the simple update, the environment tensor is approximated as a product (compare to the 1D case in fig.~\ref{fig2}b); (\textit{c}) a correlated approximation of the environment is required for the full update.
}
\label{fig3}
\end{figure*}
\end{center}

For starters, contracting PEPS is, in contrast to the efficient contraction of MPS, a ($\# P$-complete) hard computational problem~\cite{Schuch2007complexity}. 
Practical algorithms resort to approximate contractions, in which the two-dimensional network is approximated as a sequence of MPO-MPS contractions from the 
boundary~\cite{Murg2007peps,Jordan2008ipeps,Lubasch2014unifying}, 
by a coarse-graining, or tensor renormalization~\cite{Jiang2008trg-qstates,Gu2008terg} (see sec.~\ref{subsec:trg}), 
or, in the case of iPEPS, by a corner transfer matrix contraction~\cite{NishinoOkunishi1996ctmrg,Orus2009ctm}.
 In these strategies there is a trade-off between the numerical cost and the accuracy of the contraction, important to determine the 
environment of a tensor or the expectation values of observables.

Because of this, methods have been developed that gain efficiency by allowing less precise environment estimations  
for tensor updates (fig.~\ref{fig3}).
The most efficient alternative uses a so-called \emph{simple update}~\cite{Jiang2008trg-qstates} where, in order to update a pair of tensors under 
the action of a two-site gate, the environment
is approximated by a product of diagonal matrices acting on each link surrounding the pair, 
which play a role analogous to that of the Schmidt values in the TEBD procedure.
Discarding the correlations in the environment can prevent the method from
 reaching the best PEPS with fixed bond dimension in the general case~\cite{Lubasch2014unifying}, but the algorithm is still
popular, due to its efficiency and stability.
A better update can be found with
 the more expensive \emph{full update}~\cite{Jordan2008ipeps,Corboz2010fpeps},
which takes into account a more accurate correlated environment approximation. 

These approaches may improve the efficiency of the updates, 
which can be particularly useful in the case of ground state search by imaginary time evolution, where the goal is a fixed point 
of the evolution.
However, the evaluation of observables still needs to be as accurate as possible, in order to guarantee a variational result.
This yields for most of PEPS algorithms a computational cost scaling as $O(D^{10})$.


Another difference between PEPS and MPS computations is the absence of a canonical form for the former. 
As a consequence, the effective norm term appearing, for instance, in the denominator of Eq.~\ref{eq:variational}, cannot be  
reduced to the identity, and needs to be inverted to solve the local problems, which results in higher computational costs and loss of stability. 
The problem can be alleviated making use of the gauge freedom to optimize the condition of this effective matrix~\cite{Lubasch2014peps,Phien2015gauge,Evenbly2018gauge}.

Despite the higher computational challenge, and the still ongoing development of more efficient strategies,
PEPS already outperform MPS for two-dimensional problems of moderate size,
as explicitly shown in~\cite{Osorio2017pepsvsdmrg} for Heisenberg and Hubbard models.

All in all, iPEPS has been the preferred ansatz to address ground states of two-dimensional quantum problems
in this context, due to the possibility of directly addressing bulk 
properties.~\footnote{Notice that, although finite-size extrapolation from PEPS is possible, the number of tensors to determine (e.g. $L^2$ for a two-dimensional system) makes the calculations exceedingly long already for relatively small sizes.}
 {Imaginary time evolution has been the method predominantly used until very recently,} 
 due to the highly non-linear character of a variational approach in the line of \ref{subsec:classic}.
 {However, in the last few years,} 
 new strategies have been introduced for a stable and efficient variational optimization of 
iPEPS~\cite{Corboz2016variational,Vanderstraeten2016gradient}, which produces more accurate results.
A further step has been the precise solution of critical systems, with the help of extrapolations in the correlation length of finite $D$ states~\cite{Rader2018prx,Corboz2018prx,Vanhecke2021scaling}.
Plenty of impressive numerical results have been already obtained thanks to these advanced methods, 
among them the most accurate result for the Hubbard model~\cite{Zheng2017hubbard},
and the first studies of three-dimensional problems~\cite{Vlaar2021peps3D}.

When the focus is not on ground states, and similar to the MPS case discussed in section~\ref{subsec:tebd},
the (real or imaginary) time evolution techniques allow addressing multiple problems, such as 
equilibrium states at finite temperature~\cite{Czarnik2019time}, 
steady states of open systems~\cite{Kshetrimayum2017open,Kilda2021ipepo,Keever2021open}
and real time evolution~\cite{Murg2007peps, Czarnik2019time,Hubig2019tdep}.

An alternative direction has been the development and exploitation of restricted subsets of PEPS, with more favourable computational properties,
that can be suitable ansatzes for particular problems.
It is the case, for instance, of sequentially generated states~\cite{Banuls2008sgs},
 or the more general isometric PEPS~\cite{Zaletel2020isopeps}, 
or of Gaussian fermionic PEPS~\cite{Kraus2010}.

And other TNS families
that do not have an area law, or only a restricted one, can be used to study higher-dimensional systems of a certain size, as does two-dimensional DMRG. 
For instance TTN [e.g.~\cite{Tagliacozzo2009ttn2D,Magnifico2021-3d}], 
or the recently introduced augmented trees~\cite{Felser2021aTTN}.

\subsection{Symmetries} 
\label{subsec:syms}

In case the problem under study exhibits some symmetry,
taking advantage of it is not only of fundamental interest, but can also boost the performance of a numerical algorithm.
For instance, if the Hamiltonian commutes with a certain operator $[H,O]=0$, its eigenstates will have well-defined eigenvalues of $O$, and
the search can be restricted to subspaces labelled by particular quantum numbers.

In the case of quantum many-body systems, one is often interested in problems with a global symmetry of the form $U^{\otimes N}|\Psi\rangle=|\Psi\rangle$, where $U$ is a unitary transformation that acts on a single site. Particularly relevant is the case when the operation is a representation of a group $G$, namely $U=U_g,$ for some $g\in G$. 
Such Abelian symmetries were soon incorporated to the DMRG method~\cite{McCulloch2007,Schollwoeck2011},
where they became of common use, typically implemented for the conservation of particle number or total magnetization. 
Also the formalism for non-Abelian symmetries was also developed~\cite{Dukelsky1998,McCulloch2002nonAbelian}, 
albeit not so commonly used.

A general framework to handle global symmetries in higher dimensional TN was first introduced in~\cite{Singh2010sym},
with explicit formulations for Abelian~\cite{Singh2011u1,Bauer2011abelian} and non-Abelian~\cite{Singh2012su2,Weichselbaum2012} cases
following shortly.
The basic idea of these and the original DMRG constructions is to define invariant tensors, which remain unchanged when  the symmetry operation 
acts on all the indices. This requires well-defined transformation properties for each of the indices and, in particular, choosing bases 
for the virtual legs with well-defined quantum numbers, for instance
$|q \alpha \rangle$, where $q$ labels an irreducible representation of the group,\footnote{In the non-Abelian case, $q$ will actually be a composite index, 
including not only the label for the irrep, but also additional quantum numbers to account for its inner (and potentially outer) multiplicity~\cite{Bruognolo2021nonAbelian}.} 
and $\alpha$ labels the states within the same irrep.
The bond dimension of such leg will be the sum of dimensions for each $q$.
Assigning a direction to each edge in the TN, outgoing and incoming indices transform respectively with the unitary representation of the group 
and its inverse and it follows that a TNS constructed out of such invariant tensors is globally invariant.

The invariance of a tensor implies some internal structure. In the case of Abelian symmetries, the tensor can be decomposed in a 
direct sum of blocks, with the only non-vanishing ones being those for which the sum of quantum numbers of incoming indices equals 
that of outgoing ones. 
In the non-Abelian case, blocks corresponding to a suitable combination of irreps have further structure, as they can be further 
decomposed as a tensor product of one part dictated merely by the symmetry, and another one containing the free parameters of the state.
In particular, for three-legged tensors the first factor is a Clebsch-Gordan tensor. For more general tensors, a decomposition of the whole TN
in three-legged terms can be used~\cite{Singh2010sym}, or more efficient precomputation of the corresponding coefficients can be done in the algorithm~\cite{Weichselbaum2012,Hubig2018ipeps,Bruognolo2021nonAbelian}.
Notice that generic tensors (i.e. without explicit symmetry) can also be used to describe a TNS with the desired global symmetry, even producing a more compact description~\cite{Singh2013sym-vs-min}.
Using the symmetry structure of the tensors involves a more cumbersome implementation of the methods (described in detail in the previous 
references), but in exchange allows one to work with blocks which have smaller bond dimension, which reduces the computational cost of contractions at the lowest-level.

Symmetric tensors can be used to raise the global symmetry to a gauge one~\cite{Tagliacozzo2014,Zohar2015b,Haegeman2015}.
This is done through the introduction of additional link tensors (analogous to link variables in usual formulations of lattice gauge theories).

{Finally, it is worth mentioning that,} at the theoretical level, a framework has been developed to characterize MPS and PEPS in terms of the tensor 
symmetries~\cite{Schuch2010peps}, a formal approach that has produced fundamental results and continues to be an active and fruitful area of research~\cite{Cirac2021rmp}.

\subsection{Fermions}
\label{subsec:fermions}

An advantage of the TN framework with respect to other numerical methods for quantum many-body problems, 
 is the possibility to treat problems with fermionic degrees of freedom, 
of fundamental interest for condensed matter and fundamental physics.
Whereas in this case Quantum Monte Carlo methods are often obstructed by the sign problem, 
which causes the cost of convergence to increase exponentially with the system size, 
TN calculations can indistinctly treat fermionic and spin setups. 

 In one spatial dimension, fermionic modes do not pose a real problem, as they can be mapped to spins through the Jordan-Wigner transformation
 {. This maps local fermionic models onto}  local spin Hamiltonians, 
 such that both can be treated with exactly the same algorithms.
 In higher dimensions, however, a similar transformation does not maintain the locality of the model.
 An alternative that maps local fermions to local spins and would support a treatment with standard TNS algorithms was introduced in~\cite{Verstraete2005fermions},
but at the cost of introducing additional degrees of freedom doubling the size of the system.

It is however possible to define TNS directly in terms of fermionic degrees of freedom.
The explicit construction was presented by several independent, but essentially equivalent, 
proposals~\cite{Corboz2009fMERA,Kraus2010,Corboz2010fpeps,Pineda2010}. 
The fundamental idea is to work in a representation in which all spaces, virtual and physical, are fermionic, and 
have well-defined parity, i.e. the tensors are symmetric with respect to parity transformations, in the sense 
described in~\ref{subsec:syms}.
Then it is possible to encode the statistics of fermionic operators in a local way, such that the scaling of the computational cost 
with the system size is preserved.

The most intuitive formulation~\cite{Corboz2009fMERA,Corboz2010fpeps} can be visualized as an effective linear 
ordering of the fermionic modes, fixed once a graphical representation of the TNS is chosen (the order would be that 
obtained when projecting all the sites of the graph onto a line). 
Each crossing of legs in the diagram has to be accounted for, as it involves commutations of fermionic operators. This can be achieved substituting 
the crossing by a swap matrix, which introduces a negative sign with fermionic degrees of freedom with odd parity are exchanged. 
Thanks to the symmetry of the tensors, the swap matrices can be moved through the network and be absorbed into local tensors,
and the contraction can follow the same sequence as in the spin case, thus keeping the leading cost.
This formalism,
which can be combined with additional symmetries~\cite{Bruognolo2021nonAbelian},
 has already made
possible for iPEPS to beat any other computational method in some parameter regimes of the Hubbard model~\cite{Zheng2017hubbard}.

\subsection{Dynamics}
\label{subsec:dynamics}

Simulating time evolution is a crucial tool for understanding the out-of-equilibrium dynamics of quantum many-body systems,
linked to fundamental questions such as thermalization.
Together with the applicability to fermionic problems, being able to address real-time evolution is precisely one of the main advantages
of TNS as compared to Monte Carlo methods.

The TNS toolbox has several different methods to tackle these problems.
Many of them produce an approximation to the time-evolved state within the desired family [see~\cite{Paeckel2019tevol} for a recent detailed review].
The standard algorithms described in~\ref{subsec:tebd} proceed by constructing an approximation of the evolution 
operator  $U(\delta)=e^{-i \delta H}$ for a finite time step $\delta$ and applying it onto a TNS wave function.
In general, this increases the bond dimension, and it must be followed by a truncation step that reduces the tensors again.
A limitation of these methods is that they rely on approximations of the Hamiltonian exponential operator that become exceedingly costly
as the range of the interactions increases.

Krylov-based methods, instead, directly target the result of the evolution step by approximating the application of the operator on the state 
as a linear combination of Krylov vectors~\cite{GarciaRipoll2006,Wall2012njp}, instead of explicitly approximating the evolution operator 
in the full space. This in turn requires approximating the Krylov vectors themselves by TNS.
A related approach is using Chebyshev expansions of the exponential operators~\cite{Holzner2011cheMPS}.

Also the more recently proposed time dependent variational principle (TDVP)~\cite{Haegeman2011tdvp,Haegeman2016unifying}
adopts a different strategy, in which the MPS tensors are evolved such that the evolution never leaves the MPS manifold.
This is achieved by projecting the variation of the wave function, given by the rhs of the Schr\"odinger equation, onto the 
local tangent plane of the MPS.
Despite its different philosophy, TDVP algorithms for finite and infinite systems can be formulated in terms of essentially the 
same low level primitives as the traditional ones~\cite{Haegeman2016unifying}.
That is, the tensors of the ansatz can be updated according to the solution of a local evolution, in this case given by effective Hamiltonians that result from the tangent plane projection.
An advantage of this method is that it preserves conserved quantities of the evolved state, such as the norm and energy.

In the uniform MPS case, the TDVP algorithm is the first exponent of a new generation of TNS algorithms, so-called tangent-space methods~\cite{Haegeman2013post}, 
based on exploiting the geometric structure of the MPS manifold.
These increasingly popular methods have multiple applications beyond time evolution, including the variational optimization of uMPS 
or finding elementary excitations,
and have been partly adapted for PEPS [see~\cite{Vanderstraeten2018tangent} for a pedagogical overview]. More recently, generalizations for TTN and other isometric TN have been introduced~\cite{Kloss2020tdvpTree,Bauernfeind2020tdvpTTN,Hauru2021riemannian}. 

The approaches described above provide powerful algorithms to investigate the evolution of quantum systems for moderate times, 
or close to equilibrium~\cite{Paeckel2019tevol}.
However, they are still subject to the fundamental limitation mentioned in \ref{subsec:tebd}: under time evolution, entanglement can grow fast,
the bond dimension of the ansatz would need to grow exponentially with the simulated time~\cite{Osborne2006,Schuch2008},
such that after short times, the simulation becomes unfeasible, a problem that has been termed~\emph{entanglement barrier}.
But for physical problems, often the interest is not in the full description of the state, but in expectation values of local observables, 
which correspond to experimentally accessible quantities.
There a paradoxical situation takes place, since in the long time limit observables are expected to thermalize or equilibrate to values that are
well described by statistical ensembles, which can be themselves efficiently approximated by (mixed) TNS, but in most cases
the entanglement barrier makes it impossible to reach this regime following the evolution of the state.

For this reason, an active effort is being dedicated to the investigation of potentially new methods that avoid the entanglement barrier
and manage to describe the long time dynamics of local quantities. 
A first proposal was evolving operators in Heisenberg picture~\cite{Hartmann2009heisenberg} 
using a suitably adapted time evolution algorithm. Despite not completely solving the entanglement problem, such an approach 
constitutes the basis of many other strategies for dynamical quantities.
Another idea was to target the TN that represents the time-dependent local observables, and to 
approximate its contraction in the transverse direction, after folding~\cite{Banuls2009fold,Mueller-Hermes2012},
which can give access to longer times, especially when exploiting the finite propagation velocity of correlations~\cite{Frias2022lc,Lerose2022lc}.
This remains an active area of research, and
several new strategies have been proposed in the last years to focus on the local observables~\cite{White2018therm,Surace2019trading,Rakovszky2020dissip}.

\subsection{Excitations}
\label{subsec:excited}

With the variational approach {for the ground state (section~\ref{subsec:variational})} it is possible to target 
{also low excited states, }
by simply orthogonalizing the targeted state with respect to any number of 
previously computed ones~\cite{McCulloch2007,Schollwoeck2011}, an approach that 
is most useful in the case of finite systems.

A particularly useful ansatz for elementary excitations is to model them as local perturbations acting
on the vacuum. In the TNS framework, it is possible to construct well-defined momentum states of this form by suitable superpositions of 
a locally modified ground state~\cite{Oestlund1995}.
Tangent-space methods offer a way to generalize this construction that is especially 
powerful in the thermodynamic limit~\cite{Haegeman2013post,Vanderstraeten2018tangent}.
In this framework, elementary excitations are written as tangent vectors with position-dependent momentum factors,
and their energies can be optimized variationally.
Also topologically non-trivial excitations (such as domain walls) can be captured in this language.

Although low-energy excitations as the ones above are often observed to fulfill an approximate area law, 
the same is not true for generic, highly excited states.
An exception is the case of many-body localized Hamiltonians. Hence, several specific algorithms have been developed to target 
eigenstates at some high energy value $E$,
for instance
 using a \emph{shift and invert} strategy~\cite{Yu2017excited}, 
targeting the state at given energy that maximizes the overlap with a particular product state~\cite{Khemani2016exc},
or searching for the lowest eigenvalue of 
$(H-E)^2$~\cite{Lim2016exc}.

\section{FURTHER TN APPROACHES AND PERSPECTIVES}
\label{sec:other}

Other aspects of TN technologies, beyond the standard TNS tools discussed in the previous sections
offer additional ways to explore the physics of complex systems. 

\subsection{Network renormalization approaches} 
\label{subsec:trg}

Some of the earliest works in the TN literature, before the   
 quantum information perspective shaped the language for TN, 
 already pointed out the connection between many-body problems and 
tensors in the partition functions of classical spin systems~\cite{NishinoOkunishi1996ctmrg,Nishino2001twodim}.
In this approach, a TN represents exactly the partition function of a classical model (which might as well correspond to a path integral formulation of a quantum one)
and tensor contractions can be used to approximate a result.

\begin{center}
\begin{figure*}[t]
\includegraphics[width=.8\textwidth]{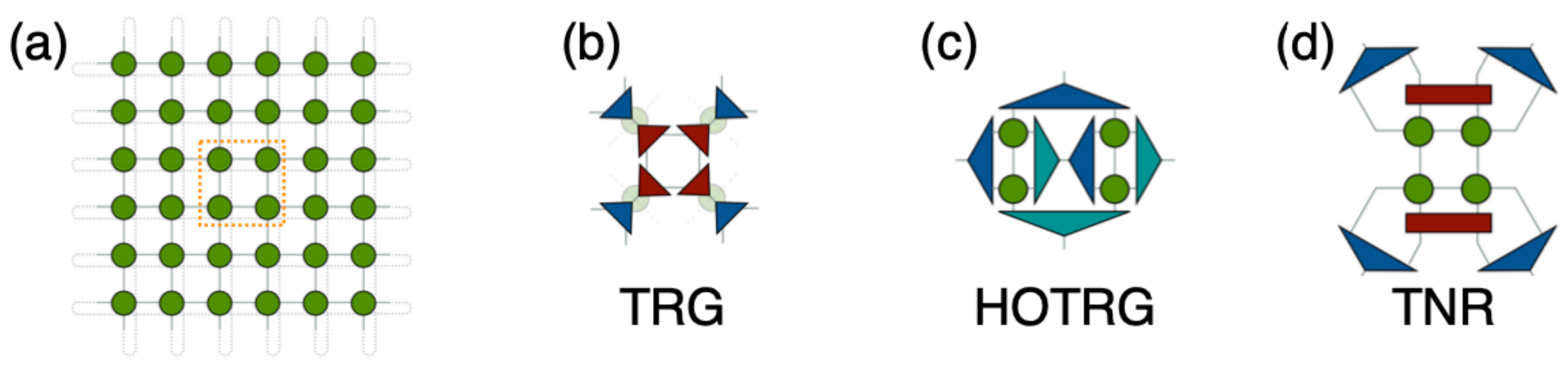}
\caption{Coarse-graining step in the simplest TRG schemes: (\textit{a}) TN representing a partition function of a classical spin model; (\textit{b}) original TRG; (\textit{c}) higher order TRG (HOTRG).
}
\label{fig4}
\end{figure*}
\end{center}

The tensor renormalization group (TRG) method introduced in~\cite{Levin2007}
is based on a block renormalization of a two-dimensional TN: in each coarse-graining step
 a local group of tensors is replaced by their approximate contraction with truncated bonds, such that 
 the size of the TN is divided by a constant (see fig.~\ref{fig4}).
 The original truncation is done by simply singular value decompositions of the tensors being contracted.
 In~\cite{Xie2009} a new strategy was introduced, called
 second renormalization group (SRG) method, where a different truncation is chosen that tries to maintain the
  fidelity of the contraction of the whole network, by taking into account the environment of the tensor that is being computed.
 A more efficient contraction and truncation strategy that can be applied to higher dimensional systems was later proposed in~\cite{Xie2012}, 
 using the higher order singular value decomposition.


 A shortcoming of the approach, already identified in~\cite{Levin2007} is that some short-range entanglement structures cannot be removed by 
the TRG coarse-graining, in particular, the corner double line (CDL) tensor.  
Several modifications have been proposed to solve this issue, 
such as the TNR (tensor network renormalization) that includes disentanglers, in the spirit of MERA, before the renormalization steps~\cite{Evenbly2015tnr}.
Other proposals have been
the iterative optimization of the tensors around a loop~\cite{Yang2015},
or different local index truncations that take care of internal correlations~\cite{Hauru2018gilt,Evenbly2018gauge}. 

 TRG approaches are also useful to contract the TN corresponding to observables for quantum states in higher 
dimensions, and can then be used as part of PEPS optimization algorithms~\cite{Jiang2008trg-qstates,Gu2008terg}
 (see sec.~\ref{subsec:peps}).
A related topic is the treatment of fermionic problems in TRG approaches.
In \cite{Gu2010grassmann} it was shown that wave functions and expectation values of many-body fermionic (but also bosonic) systems
could be expressed and contracted as a Grassmann tensor network, in which tensor components are given in terms of Grassmann variables,
and for which a suitable TRG approach can be defined.
A compact ansatz of this form, together with algorithms to renormalize the network and to evolve the tensors
were presented in~\cite{Gu2013gtrg}, and have been used, for instance, 
to study discretized field theories with fermionic degrees of freedom  [see references in~\cite{Banuls2020ropp,Meurice2020review}].

\subsection{Connections to other techniques} 
\label{subsec:connections}

Exploring the potential connections between TN methods and other techniques is an exciting possibility 
that, on the one hand, can result in new or improved algorithms and, on the other, opens the door to treating new problems with TN methods,
as the following examples illustrate.

\begin{itemize}
\item{Monte Carlo algorithms.} 
Monte Carlo sampling can be used to speed up TN contractions, and variationally optimize TNS 
parameters~\cite{Sandvik2007mc,Wang2011mctns}. 
With a complementary perspective, TN contractions can be employed to directly sample configurations from the partition function~\cite{Ueda2005snapshot,Ferris2012perfect,Rams2021sampling},
but also to define a Markov chain with collective updates~\cite{Frias2021tnmh}.
\item{Machine Learning.} 
The connections between TN and machine learning drive some of the most recent developments,
including the use of TNS models for machine learning tasks~\cite{Stoudenmire2016nips,Han2018unsupervised},
and also importing numerical tools, such as automatic differentiation, into TN algorithms~\cite{Liao2019differentiable}. 
\item{Field theory.} 
The interplay between TNS and quantum field theory is another decidedly active area, which has 
produced accurate numerical results for lattice gauge theories~\cite{Banuls2020ropp,Meurice2020review}, but also
motivates formal developments, such as gauge symmetric (see sec.~\ref{subsec:syms}) and continuous~\cite{Verstraete2010cMPS,Haegeman2013cMERA} formulations of TNS. 
\end{itemize}

\section{OUTLOOK}
\label{sec:outlook}

The field of tensor networks has grown impressively in the last decade and
remains a vibrant research area.
Current TN research moves forward in different directions.
A rather formal approach explores the mathematical aspects of
these ansatzes.
With a more applied perspective, significant effort is being devoted to the development of
numerical TN methods, a multifaceted enterprise, some of whose spotlights have been highlighted in the previous pages.
And the field continues to uncover synergies with seemingly remote topics,
and to develop in new and creative ways.
All these directions are  likely to produce exciting results in the coming years, 
maybe finding useful TNS subfamilies, 
improving the efficiency of high-dimensional or dynamical calculations, or 
bridging the gaps between formal and numerical developments.

At the same time, mature TN algorithms are well established as competitive computational methods 
for the study of many-body problems.
These algorithms, reviewed in the first part of this article, 
make it easy for the newcomer to try TN for an existing problem,
 and simultaneously serve as a platform for the more specialized researcher 
  to experiment with new algorithms or to draw new connections between TN and other disciplines.

\section*{ACKNOWLEDGMENTS}
I am deeply grateful to E. Carmona, P. Emonts, M. Fr{\'{\i}}as-P{\'e}rez and T. Nishino for their critical reading and constructive comments on an earlier version of this article.
This work was partly funded by the Deutsche Forschungsgemeinschaft (DFG, German Research Foundation) under Germany's 
Excellence Strategy -- EXC-2111 -- 390814868.

%

\bibliographystyle{ar-style4}
\bibliography{TNS_cmp}

\section*{RELATED RESOURCES}

\begin{enumerate}
\item{T. Nishino, The DMRG Homepage, http://quattro.phys.sci.kobe-u.ac.jp/dmrg.html}
\item{E. M. Stoudenmire, The Tensor Network, https://tensornetwork.org/}
\item{M. Fishman, E. M. Stoudenmire, S.  R.White, An open source TNS library, https://itensor.org/}
\item{J. Hauschild, F. Pollmann, A tensor library for Python (TeNPy), https://github.com/tenpy/tenpy}
\item{G. Evenbly, Tensors.net, https://www.tensors.net/}
\end{enumerate}

\end{document}